\documentclass[conference]{IEEEtran}

\usepackage{cite}
\usepackage{graphicx}
\usepackage{caption}
\usepackage{subcaption}

\usepackage[cmex10]{amsmath}
\usepackage{algorithmic}
\usepackage{array}
\usepackage[font=footnotesize]{subfig}
\usepackage{url}
\usepackage{fontenc}
\usepackage[utf8]{inputenc}
\usepackage{mathtools}
\usepackage{tabularx}
\usepackage{algorithmic}
\newsavebox{\ieeealgbox}
\newenvironment{boxedalgorithmic}
  {\begin{lrbox}{\ieeealgbox}
   \begin{minipage}{\dimexpr\columnwidth-2\fboxsep-2\fboxrule}
   \begin{algorithmic}}
  {\end{algorithmic}
   \end{minipage}
   \end{lrbox}\noindent\fbox{\usebox{\ieeealgbox}}}
\newcolumntype{M}[1]{>{\arraybackslash}m{#1}}
\newcolumntype{N}{@{}m{0pt}@{}}
\begin{document}
\title{Opportunistic Spectrum Allocation for Max-Min Rate in NC-OFDMA}

% author names and affiliations use a multiple column layout for up to
% three different affiliations
\author{\IEEEauthorblockN{Ratnesh Kumbhkar, Tejashri Kuber, Narayan B. Mandayam, Ivan Seskar} \IEEEauthorblockA{WINLAB, Rutgers
University\\ 671 Route 1, North Brunswick NJ 08902\\ Email:
\{ratnesh, tkuber, narayan, seskar\}@winlab.rutgers.edu}}
\IEEEoverridecommandlockouts
%\IEEEpubid{\makebox[\columnwidth]{978-1-4244-8953-4/11/\$26.00~\copyright~2015 IEEE \hfill} \hspace{\columnsep}\makebox[\columnwidth]{ }} 
%..
\maketitle

%%%%%%%%%%%%%%%%%%%%%%%%%%%%%%%%%%%%%%%%%%%%%%%%%%%%%%%%%%%%%%%%%%%%%%%%%
%%%%%%%%%%%%%%%%%%%%%%%%%%%%%%%%%%%%%%%%%%%%%%%%%%%%%%%%%%%%%%%%%%%%%%%%%
\begin{abstract}
%\boldmath
%Cognitive radios (CRs) are typically deployed to make better use of under-utilized licensed spectrum bands. Operating under such environments requires techniques for efficent use of non-contiguous spectrum bands. 
We envision a scenario of opportunistic spectrum access among multiple links when the available spectrum is not contiguous due to the presence of external interference sources. Non-contiguous Orthogonal Frequency Division Multiplexing (NC-OFDM) is a promising technique to utilize such disjoint frequency bands in an efficient manner. In this paper we study the problem of fair spectrum allocation across multiple NC-OFDM-enabled point-to-point cognitive radio links under certain practical considerations that arise from such non-contiguous access. When using NC-OFDMA, the channels allocated to a cognitive link are spread across several disjoint frequency bands leading to a large \textit{spectral span} for that link. Increased spectral span requires higher sampling rates, leading to increased power consumption in the ADC/DAC of the transmit/receive nodes. 
%It is thus pertinent to pursue techniques for spectrum allocation that also account for the spectral span. 
In this context, this paper proposes a spectrum allocation that maximizes the minimum rate achieved by the cognitive radio links, under a constraint on the maximum permissible spectral span. Under constant transmit powers and orthogonal spectrum allocation, such an optimization is a mixed-integer linear program and can be solved efficiently. 
%using readily available solvers. 
There exists a clear trade-off between the max-min rate achieved and the maximum permissible spectral span. The spectral allocation obtained from the proposed optimization framework is shown to be close to the trade-off boundary, thus showing the effectiveness of the proposed technique. We find that it is possible to limit the spectrum span without incurring a significant penalty on the max-min rate under different interference environments. We also discuss an experimental evaluation of the techniques developed here using the Universal Software Radio Peripheral (USRP) enabled ORBIT radio network testbed. 
\end{abstract}

%%%%%%%%%%%%%%%%%%%%%%%%%%%%%%%%%%%%%%%%%%%%%%%%%%%%%%%%%%%%%%%%%%%%%%%%%
%%%%%%%%%%%%%%%%%%%%%%%%%%%%%%%%%%%%%%%%%%%%%%%%%%%%%%%%%%%%%%%%%%%%%%%%%
\section{Introduction}
As the number of devices using wireless spectrum has increased, availability of usable spectrum for the licensed devices is a concern. Cognitive radio (CR) plays an important role in addressing this problem with dynamic spectrum access. In the past few years there has been a large amount of research on addressing different aspects of cognitive radios (e.g. \cite{ mitola99cognet, pricognet, haykin05cognetbrain, thomas05cognet,  ileri08dsamodel, liang08sense, srini07cognetdsa, Cesana11routing, cheng07joint, NazmulWiOpt}). Orthogonal frequency division multiplexing (OFDM)  has been suggested as one of the candidates for dynamic spectrum access in CRs due to its flexible and efficient use of the spectrum \cite{weiss04}. Non-contiguous OFDM (NC-OFDM) is a method of transmission where some of the subcarriers in OFDM are nulled and  only the remaining subcarriers are used for transmission \cite{rajbanshi06ncofdmrx, UCSB1, UCSB2, NCOFDM_Implementation}. Since available unused spectrum is generally non-contiguous,  usage of  NC-OFDM  results in better spectrum utilization. Further, since NC-OFDM allows the CRs to access the unused spectrum without interfering with the licensed users, it also complies with the broader objective  that primary users of the spectrum need not consider the presence of CRs and can be completely oblivious to them.
Techniques for efficient implementation of the DFT operation for NC-OFDM when multiple subcarriers are nulled are also available. \cite{rajbanshi06ncofdmrx}. However, one main drawback of NC-OFDM is that it suffers from high out-of-band radiation due to the high sidelobes of its modulated subcarriers, which can potentially affect the performance of licensed users, or other CRs in the unlicensed band. Several techniques to address this issue have been proposed and we briefly touch upon these issues in the latter part of this paper. 

A significant concern when using NC-OFDMA is that the cognitive links are allocated disjoint frequency bands that lead to an increased spectral span of a cognitive link. The spectral span is defined as the difference between the frequencies of the extreme channels allocated to a cognitive link. Increase in the spectral span leads to higher sampling rates that in turn lead to an increase in the power consumption at the transmit/receive nodes.
Usually, the transmit power requirements of a transceiver system have dominated the total power consumption. However, the ADC/DAC power consumption can become comparable or even significantly larger than the transmit power consumption when the sampling rates become very large \cite{Nazmul_NCOFDM1}. Therefore, it is important to impose a reasonable limit on the spectrum span.
%$<$Need some discussion here on system power consumption vs. transmit power consumption in CRs$>$.

In this paper, we consider the problem of  spectrum allocation across multiple point-to-point cognitive links between  NC-OFDM-enabled transceivers in the presence of interference from the primary users. The main goal is to achieve a fair spectrum allocation that maximizes the minimum data rate across these cognitive links while limiting the spectral span. Towards this goal, we propose an optimization framework to maximize the minimum rate, subject to the constraint that spectrum span is not too large. Under constant transmit powers and orthogonal spectrum allocation, such an optimization is a mixed-integer linear program and can be solved efficiently using readily available solvers. Simulation results show a trade-off between the max-min rate and spectrum span. In our simulations, we also show improvement in data rate based on spectrum allocation obtained from solving the optimization problem in presence of interference. We also implement the NC-OFDM system using USRP \cite{ettus} radios with GNU Radio software platform on ORBIT testbed \cite{ray06cognet}. GNU Radio is a free and open-source software development toolkit that provides signal processing blocks to implement software radios \cite{gnuradio}.

The remainder of the paper is organized as follows. The related work is presented in section \ref{sec:relwork}. In section \ref{sec:sys} we present  our system model with various channel and allocation constraints and as well as the  problem formulation. In section \ref{sec:sim} we present our simulation setup and  simulation results. The experimental setup on the  ORBIT testbed and corresponding results are shown in section \ref{sec:exp} and we  conclude in section \ref{sec:conclude}.
%%%%%%%%%%%%%%%%%%%%%%%%%%%%%%%%%%%%%%%%%%%%%%%%%%%%%%%%%%%%%%%%%%%%%%%%%
%%%%%%%%%%%%%%%%%%%%%%%%%%%%%%%%%%%%%%%%%%%%%%%%%%%%%%%%%%%%%%%%%%%%%%%%%
\section{Related work}
\label{sec:relwork}
While optimizing communication links for total transmit power is a well studied area, considerations for total system power consumption is an area of active research. The authors of \cite{Li11energysurvey} consider the effect of system power for energy efficient wireless communications. Modulation schemes optimized for system power consumption are studied in \cite{shug05energyopt}, while the authors in \cite{grover11syspower} present a communication-theoretic view of system power consumption. System power constraints specifically related to NC-OFDM are studied in in \cite{jia11cap, cao10dsa}, where it is shown that the maximum spectral span is limited by ADC/DAC \cite{jia11cap} and that the requirement of a guard band affects the overall system throughput. Nazmul et. al \cite{Nazmul_NCOFDM1} characterize the trade-off between the system power and spectrum span from a cross-layer perspective in a multi-hop network. The authors in \cite{zhang10grcol} provide a graph coloring method for  spectrum allocation with the goal of providing equal rates to each user. To the best of our knowledge, previous works have not considered fair spectrum allocation with system power considerations for NC-OFDM. Our work focuses on the opportunistic spectrum allocation to maximize the max-min rate while limiting the spectral span of the NC-OFDM-enabled cognitive radio links.

%%%%%%%%%%%%%%%%%%%%%%%%%%%%%%%%%%%%%%%%%%%%%%%%%%%%%%%%%%%%%%%%%%%%%%%%%
%%%%%%%%%%%%%%%%%%%%%%%%%%%%%%%%%%%%%%%%%%%%%%%%%%%%%%%%%%%%%%%%%%%%%%%%%
\section{System Model}
\label{sec:sys}
We consider  a network of $N$ point-to-point links  that use NC-OFDM for communication. The set of $N$ links in this model is represented by $\mathcal{N}$. These links have access to $M$ channels, represented by the  set $\mathcal{M} = \{1,2,\ldots,M\}$, with each channel having a bandwidth of $W$ Hz. We assume that each channel supports $t$ OFDM subcarriers.  Transceivers in these links can be dynamically programmed to use different sets of channels.  The distance between the transmitter and the receiver in link $l$ is denoted as $d_l$. The channel gain  for link $l$ on the $m$th channel is represented as $g_l^m$. The link gain encompasses antenna gain, coding gain and fading. The received power at the receiver of link $l$ on the channel $m$ is given by $g_l^m p_l^m$.  We assume that each channel experiences flat fading and that there is no correlation between any two channels. 
We denote the $N\times M$ channel allocation matrix by  $\textbf{A}$. Elements of matrix $\textbf{A}$ can either be 1 or 0. The $i$th row of $\textbf{A}$ represents the channel allocation vector for the $i$th link. Elements of $\textbf{A}$ are defined  as follows\\
%%%%%%%%%%%%%%%%%%%%%%%%%%%%%%%%%%%%%%%%%%%%%
\begin{equation*}
	a_{lm}=\begin{cases}
    1, \;\;\;\text{link $l$ is scheduled on channel $m\in\mathcal{M}$}\\
    0,\;\;\;\text{otherwise.}
    \end{cases}
\end{equation*}
%%%%%%%%%%%%%%%%%%%%%%%%%%%%%%%%%%%%%%%%%%%%%
%%%%%%%%%%%%%%%%%%%%%%%%%%%%%%%%%%%%%%%%%%%%%
\begin{table}[b]
\caption{List of notations}
\centering
\begin{tabular}{|M{0.7cm}|M{0.39\textwidth}|N}
  \hline
  $\mathcal{N}$ & Set of links &\\[3pt]
  \hline
  $N$ & Number of links &\\[3pt]
  \hline
  $\mathcal{M}$ & Set of total available channels &\\[3pt]
  \hline
  $M$ & Number of total available channels &\\[3pt]
  \hline
  $\textbf{A}$ & Resource allocation matrix of size $N\times M$ &\\[3pt]
  \hline
  $a_{lm}$ & Allocation indicator variable for link $l$ and channel $m$&\\[3pt]
  \hline
  $\textbf{U}$ & Interference matrix of size $N\times M$ &\\[3pt]
  \hline
  $u_{lm}$ & Interference for link $l$ on channel $m$&\\[3pt]
  \hline
  $g_l^m$ & Channel gain for link $l$ using channel $m$&\\[3pt]
  \hline
  $c_l^m$ & Channel capacity for link $l$ using channel $m$&\\[3pt]
  \hline
  $r_l^m$ & Data rate for link $l$ using channel $m$&\\[3pt]
  \hline
  $r_l$ & Total data rate for link $l$ &\\[3pt]
  \hline
  $d_l$ & Distance between transmitter and receiver for link $l$ &\\[3pt]
  \hline
  $W$ & Bandwidth of each channel&\\[3pt]
  \hline
  $N_0$ & Noise spectral density&\\[3pt]
  \hline
\end{tabular}
\label{table:param}
\end{table}
%%%%%%%%%%%%%%%%%%%%%%%%%%%%%%%%%%%%%%%%%%%%%
%%%%%%%%%%%%%%%%%%%%%%%%%%%%%%%%%%%%%%%%%%%%%
\begin{figure}
        \centering
        \begin{subfigure}[b]{0.24\textwidth}
                \includegraphics[width=\textwidth]{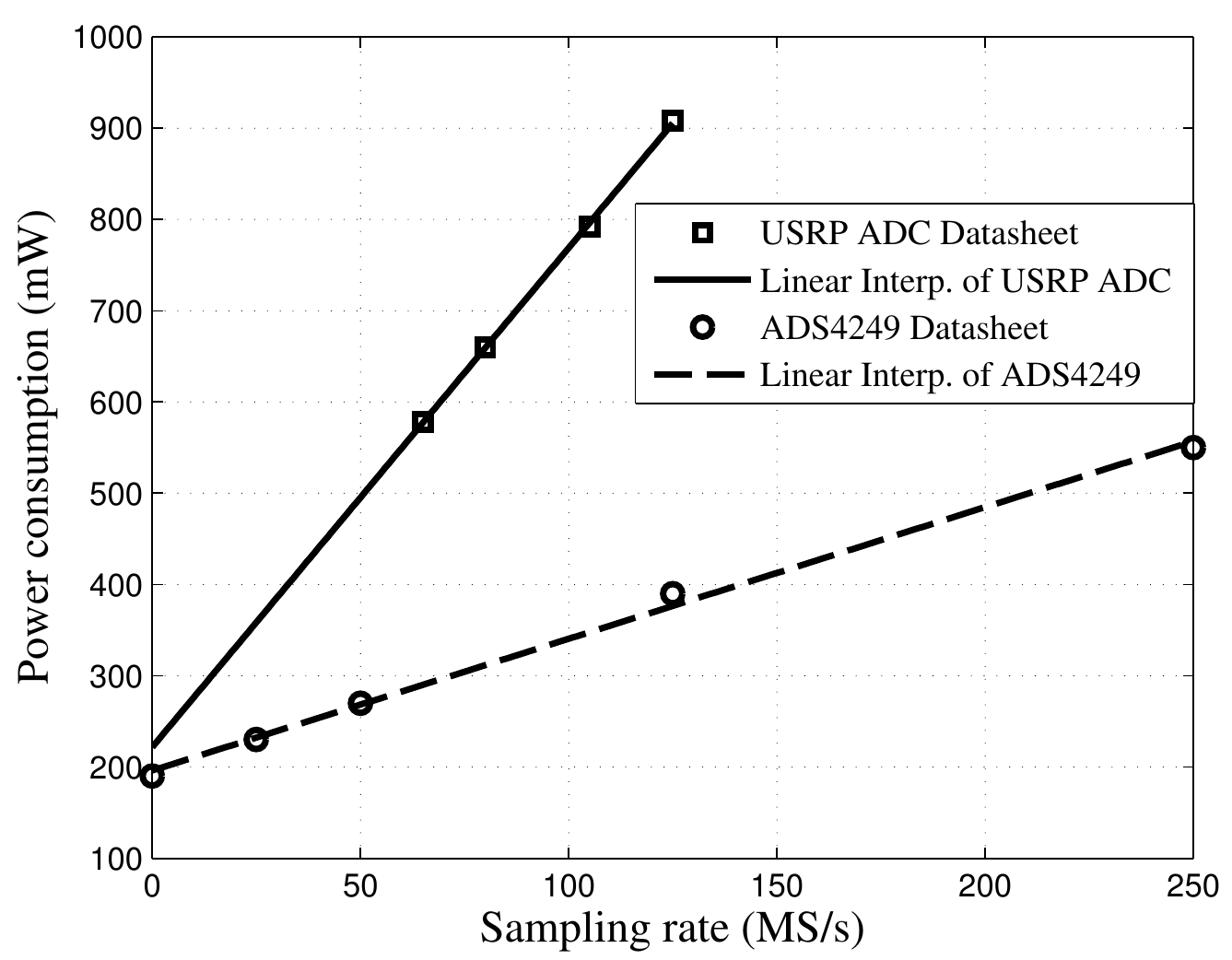}
                \caption{ADC}
                \label{fig:adc}
        \end{subfigure}%
        \begin{subfigure}[b]{0.24\textwidth}
                \includegraphics[width=\textwidth]{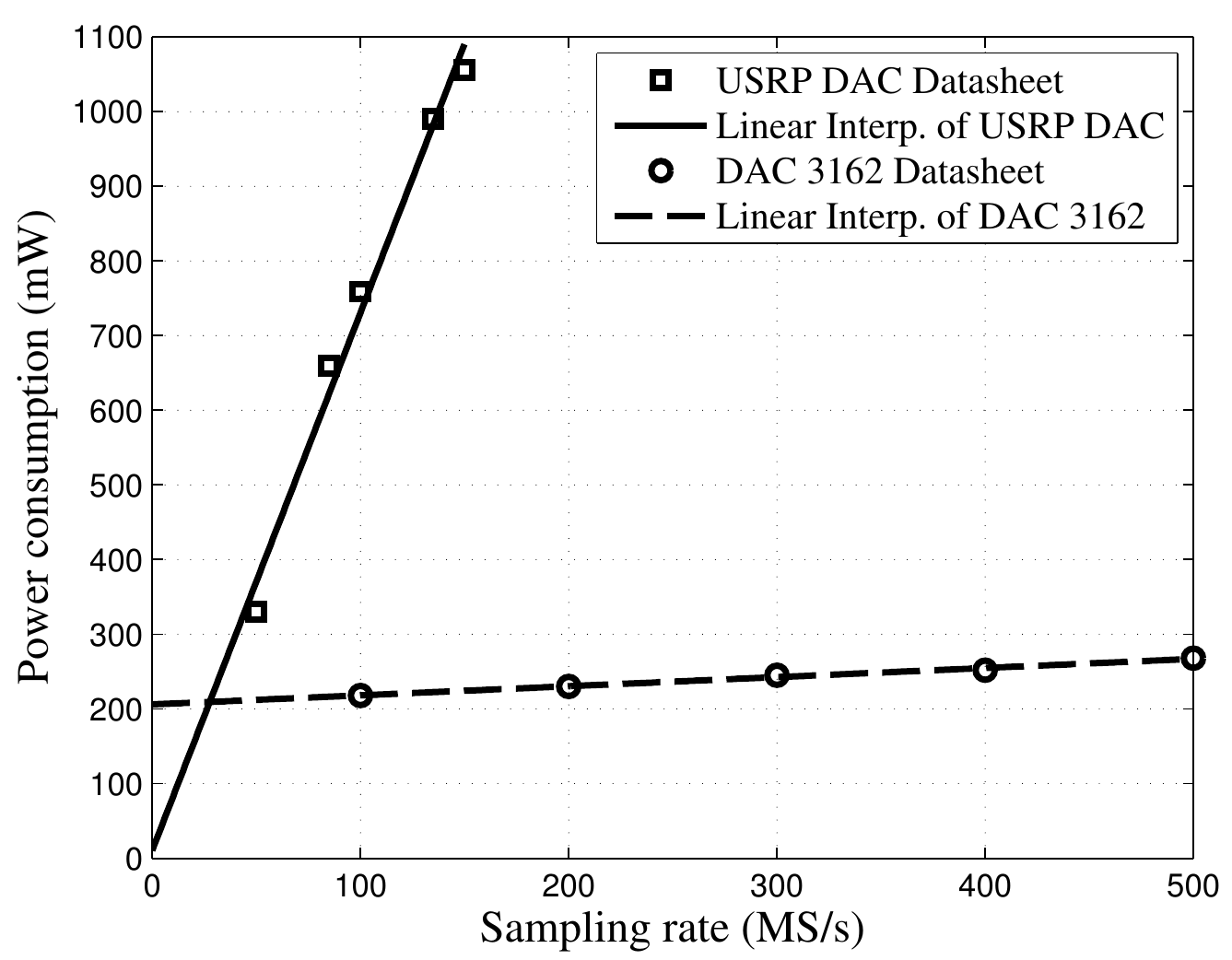}
                \caption{DAC}
                \label{fig:dac}
        \end{subfigure}
        \caption{Power consumption in the ADC and DAC of USRP  \cite{Nazmul_NCOFDM1, ad9777, ads62, dac3162, ads4249}}\label{fig:syspower}.
\end{figure}
%%%%%%%%%%%%%%%%%%%%%%%%%%%%%%%%%%%%%%%%%%%%%
We assume that all $N$ links in our model can potentially interfere each other, and hence restrict ourselves to disjoint or orthogonal allocation of the available channels. Thus,
%%%%%%%%%%%%%%%%%%%%%%%%%%%%%%%%%%%%%%%%%%%%%
\begin{align}
  \sum_{l=1}^N a_{lm} &\leq 1,\;\;\; \forall\; m \in \mathcal{M}. %\nonumber
\end{align}
%%%%%%%%%%%%%%%%%%%%%%%%%%%%%%%%%%%%%%%%%%%%%
As discussed in the previous section, the total spectrum span of a cognitive link affects the sampling rate and hence the system power. Fig. \ref{fig:syspower} reproduced from  \cite{Nazmul_NCOFDM1}, shows the power consumption in the ADCs and  DACs that are typically used in USRP radios as a function of sampling rate. Higher bandwidth usage results in higher sampling rate, and this  increases system power consumption in the ADC and DAC. Therefore, it becomes important to keep the overall spread of frequencies over which the channels are allocated to a link to a reasonably small value. We define the spectral span $B_l$ for a link $l$ as difference in frequency between channels with smallest and largest index. For a link $l$, this can be written as
%%%%%%%%%%%%%%%%%%%%%%%%%%%%%%%%%%%%%%%%%%%%%
\begin{align}
  B_l=&\Big(\max_{m \in \mathcal{M}}(m\cdot a_{lm})\nonumber\\ 
  &- \min_{m \in \mathcal{M}}(m\cdot a_{lm}+M(1-a_{lm}))+1\Big)\cdot W\;\;\;  %\nonumber
\end{align}
%%%%%%%%%%%%%%%%%%%%%%%%%%%%%%%%%%%%%%%%%%%%%
% where
% \begin{align}
%   E_l&= \{m\cdot a_{lm}\}_{m\in \mathcal{M}}\;\;\; \forall\; l \in \mathcal{N}.\, \nonumber
% \end{align}
We define a  threshold $b$ for the spectral span such that
%%%%%%%%%%%%%%%%%%%%%%%%%%%%%%%%%%%%%%%%%%%%%
\begin{align}
  B_l&\leq b\cdot W\;\;\; \forall\; l \in \mathcal{N}\, %\nonumber
  \end{align}
%%%%%%%%%%%%%%%%%%%%%%%%%%%%%%%%%%%%%%%%%%%%%
  and
%%%%%%%%%%%%%%%%%%%%%%%%%%%%%%%%%%%%%%%%%%%%%
  \begin{align}
    \left\lceil \frac{M}{N}\right\rceil &\leq b \leq  M.%\nonumber
\end{align}
%%%%%%%%%%%%%%%%%%%%%%%%%%%%%%%%%%%%%%%%%%%%%
The lower bound for $b$ results from the fact that our objective is to get the max-min rate for all links while using all available spectrum. Any lower value of $b$ will result in inefficient usage of spectrum.

We also assume that each of the $N$ links  experience  interference from  a different set of out-of-network interfering nodes and we have no control over these interfering nodes. The matrix $\textbf{U}$ of size $N\times M$ represents the interference power observed by the links, where element $u_{lm}$ of this matrix represents the interference power observed by the receiver of link $l$ on channel $m$.
The transmit power used by link $l$ on channel $m$ is represented by $p_l^m$ and it is  kept at a constant value throughout this paper. Based on the power allocation and the channel gain, the signal-to-interference-plus-noise ratio (SINR) on channel $m$ for the receiver of link $l$ is defined as 
%%%%%%%%%%%%%%%%%%%%%%%%%%%%%%%%%%%%%%%%%%%%%
\begin{align}
  s_{l}^m &= \frac{p_{l}^m g_{l}^m}{N_0 W+u_{lm}}\;\;\; \forall\; l \in \mathcal{N},\; m \in \mathcal{M}\, %\nonumber
\end{align}
%%%%%%%%%%%%%%%%%%%%%%%%%%%%%%%%%%%%%%%%%%%%%
where $N_0$ is the noise power spectral density. Since we have assumed disjoint channel allocation, we do not consider the interference from other links while calculating the SINR. When the channel $m$ is allocated to link $l$, the data rate for link $l$ on channel $m$ is bounded by the capacity, which is defined as
%%%%%%%%%%%%%%%%%%%%%%%%%%%%%%%%%%%%%%%%%%%%%
\begin{align}
  c_{l}^m &= W \log_2 (1+s_{l}^m) \;\;\; \forall\; l \in \mathcal{N},\; m \in \mathcal{M}. %\nonumber
\end{align}
%%%%%%%%%%%%%%%%%%%%%%%%%%%%%%%%%%%%%%%%%%%%%
Depending on whether this channel is allocated to this link or not, the rate $r_l^m$ achieved by link $l$ on this channel satisfies
%%%%%%%%%%%%%%%%%%%%%%%%%%%%%%%%%%%%%%%%%%%%%
  \begin{align}
  r_{l}^m &\leq c_{l}^m a_{lm}. %\nonumber
\end{align}
Thus, the total data rate obtained for link $l$ is denoted as $r_l$, which is given as 
\begin{align}
  r_l &= \sum_{m=1}^M r_{l}^m\;\;\; \forall\; l \in \mathcal{N}.\, %\nonumber
\end{align}
%%%%%%%%%%%%%%%%%%%%%%%%%%%%%%%%%%%%%%%%%%%%%

%Here we assume that the channel allocation is decided by an outside node which behaves as a controller, and decides the initial allocation of channels based on spectrum sensing and allocates the channels after the introduction of interference by implementing the algorithm we describe.The parameters used for these links are given in table \ref{table:param}.

%The NC-OFDM system in GNU Radio is 

%%%%%%%%%%%%%%%%%%%%%%%%%%%%%%%%%%%%%%%%%%%%%%%%%%%%%%%%%%%%%%%%%%%%%%%%%
\subsection{Problem formulation} 
\label{sec:problem}
The objective of this paper is to obtain a fair spectrum allocation across all the cognitive links in the system such that it maximizes the minimum data rate among all the links. The spectrum is allocated in an orthogonal manner and the resulting span is restricted to be within a given threshold, so as to limit the overall system power consumption.

To achieve the stated objectives, we formulate an optimization problem to maximize the minimum data rate while restricting the spectrum span to be below a threshold of $b$ channels for each link. Such an optimization problem can be written as follows
%%%%%%%%%%%%%%%%%%%%%%%%%%%%%%%%%%%%%%%%%%%%%
\begin{equation*}
\begin{aligned}
&{\text{maximize}}\;\;\; \min_{l\in \mathcal{N}}{r_l} \\
&\text{subject to :}\\
& B_l\leq b\cdot W\;\;\;\;&\forall \; l\in \mathcal{N},\\
%& B_l=(\max_m(E_l) - \min_m(E_l+M(1-a_{lm}))&+1)\cdot W,\\
%& E_l= \{m\cdot a_{lm}\}_{m\in \mathcal{M}}\;\;\; &\forall\; l \in \mathcal{N}\\
%& p_l^m = P \;\;\;\;&\forall \; l\in \mathcal{N},\\
%&c_l^m = W \log_2(1+s_l^m) \;\;\;\; &\forall \, l \in \mathcal{N},\\  
%&s_l^m = \frac{p_l^m g_l^m}{N_0 W + {u}_{lm}}\\
&r_l^m \leq c_l^m . a_{lm} \;\;\;\;&\forall \; l\in \mathcal{N},\;\forall\; m \in \mathcal{M},\\
&r_l = \sum_{m=1}^M r_{l}^m \;\;\; &\forall\; l \in \mathcal{N},\\
&\sum_{l=1}^N a_{lm} \leq 1,\;\;\; &\forall\; m \in \mathcal{M},\\
&r_l^m \geq 0,\ a_{lm}\in \{0,1\}\;\;\;\;&\forall \; l\in \mathcal{N},\;\forall\; m \in \mathcal{M}.
\end{aligned}
\end{equation*}
%%%%%%%%%%%%%%%%%%%%%%%%%%%%%%%%%%%%%%%%%%%%%
We note that the above problem formulation is a mixed-integer linear program. Note that $c_l^m$ is a constant in the above formulation as the transmit powers are held constant and spectrum is allocated orthogonally. Maximizing the minimum rate and restricting the spectral span are two competing objectives. Allowing a higher value of $b$ provides the opportunity to allocate the channels over a wider range of possibilities which might result in higher data rate, but this increases  the system power consumption. On the other hand keeping the spectrum span threshold too small eliminates these allocation opportunities. We analyze this trade-off in detail in the next section.

%%%%%%%%%%%%%%%%%%%%%%%%%%%%%%%%%%%%%%%%%%%%%
\begin{figure}[!t]
  \centering
    \includegraphics[width=0.38\textwidth]{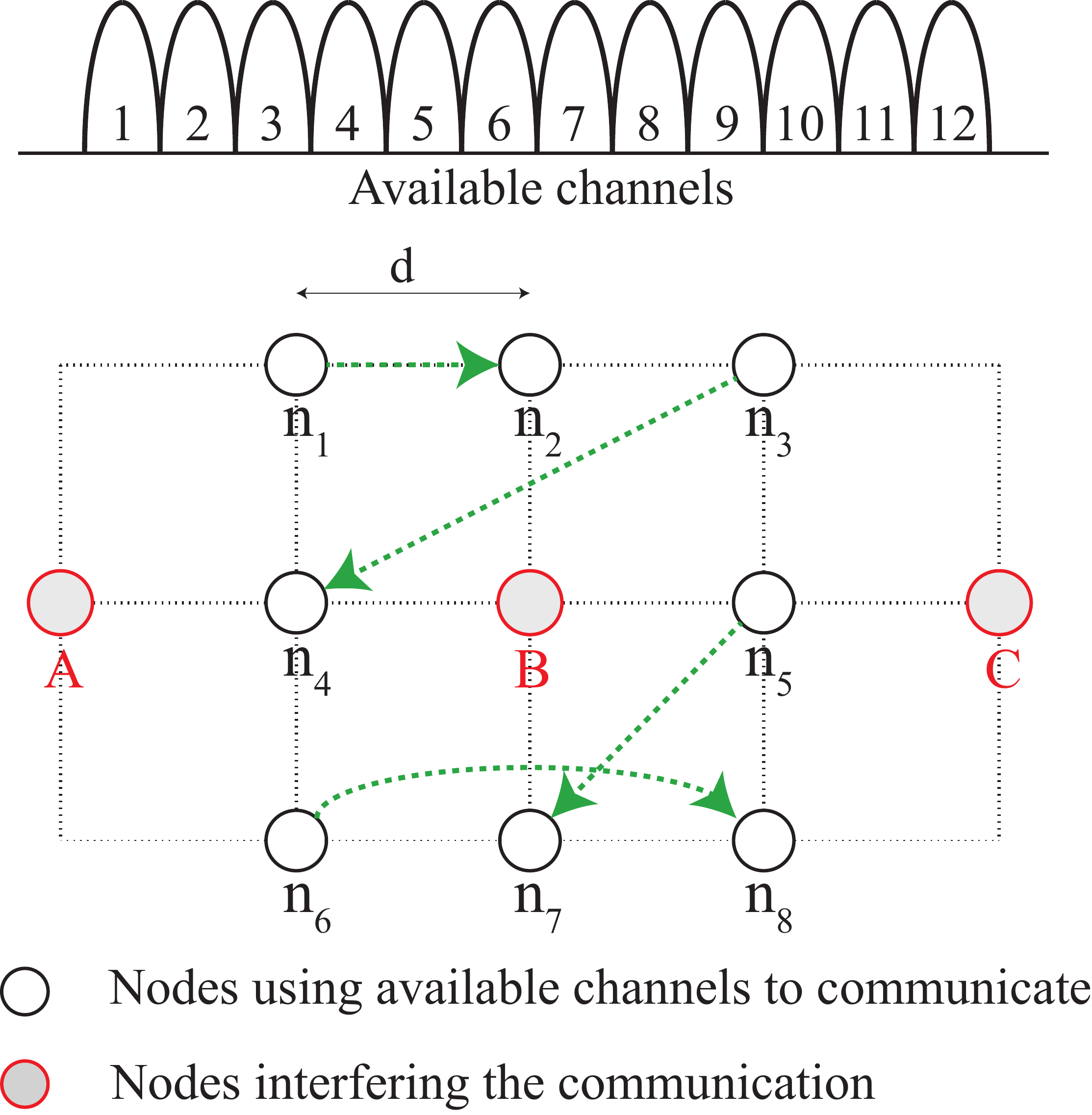}
    \caption{Available channels and network topology used in the simulation.}
  \label{fig:toposim}
\end{figure}
%%%%%%%%%%%%%%%%%%%%%%%%%%%%%%%%%%%%%%%%%%%%%

%%%%%%%%%%%%%%%%%%%%%%%%%%%%%%%%%%%%%%%%%%%%%
\begin{table}[t]
\centering
\begin{tabular}{ c|c|c c } 
\hline
Link & Nodes & Length \\
\hline
$L_1$ & $n_1 \rightarrow n_2$ & $d$\\ 
$L_2$ & $n_3 \rightarrow n_4$ & $\sqrt{5}d$\\ 
$L_3$ & $n_5 \rightarrow n_7$ & $\sqrt{2}d$\\ 
$L_4$ & $n_6 \rightarrow n_8$ & $2d$\\ 
\hline
\end{tabular}
\caption{Links in the network used for the simulation.}
\label{table:link}
\end{table}
%%%%%%%%%%%%%%%%%%%%%%%%%%%%%%%%%%%%%%%%%%%%%

%%%%%%%%%%%%%%%%%%%%%%%%%%%%%%%%%%%%%%%%%%%%%%%%%%%%%%%%%%%%%%%%%%%%%%%%%
%%%%%%%%%%%%%%%%%%%%%%%%%%%%%%%%%%%%%%%%%%%%%%%%%%%%%%%%%%%%%%%%%%%%%%%%%
\section{Simulation results}
\label{sec:sim}
The mixed-integer linear program formulated in the previous section cab be solved using the MOSEK solver with CVX in MATLAB \cite{cvx, mosek}.  MOSEK solves the mixed-integer program using the branch-and-bound method, which is known to have an exponential complexity. The output of such an optimization generates a list of channel allocations for each link along with the rates achieved in each of them.
To analyze the effectiveness of proposed approach, we use the  topology shown in Fig. \ref{fig:toposim}. We assume that there are 12  channels  available for communication, with each channel having a bandwidth of  100 KHz. As shown in  Fig. \ref{fig:toposim}, the nodes named $n_1$ to $n_8$ use these channels in a cognitive manner without affecting the primary transmitters $A$, $B$ and $C$. Nodes $n_1$, $n_3$, $n_5$ and $n_6$ are assumed to be transmitters, transmitting to nodes $n_2$, $n_4$, $n_7$ and $n_8$ respectively as shown in table \ref{table:link}. In our simulation, we assume that grid spacing is $d=1$m and the transmission power is $0.1$mW. The noise power is calculated from the thermal noise power density assuming that our system operates at a temperature of $T=300$K. For such a system, the parameters corresponding the system model are given as follows

%%%%%%%%%%%%%%%%%%%%%%%%%%%%%%%%%%%%%%%%%%%%%
\begin{align}
  \mathcal{N} &= \{L_1,L_2,L_3,L_4\},\, \nonumber\\
  \mathcal{M} &= \{1,2,3,4,5,6,7,8,9,10,11,12\}, \nonumber\\
  W&=100\mbox{KHz}, \nonumber\\
  N_0&=kT, \;\; k=\mbox{Boltzman constant}.\nonumber
\end{align}
%%%%%%%%%%%%%%%%%%%%%%%%%%%%%%%%%%%%%%%%%%%%%

%%%%%%%%%%%%%%%%%%%%%%%%%%%%%%%%%%%%%%%%%%%%%
\begin{figure}[!t] 
  \centering
    \includegraphics[width=0.3\textwidth]{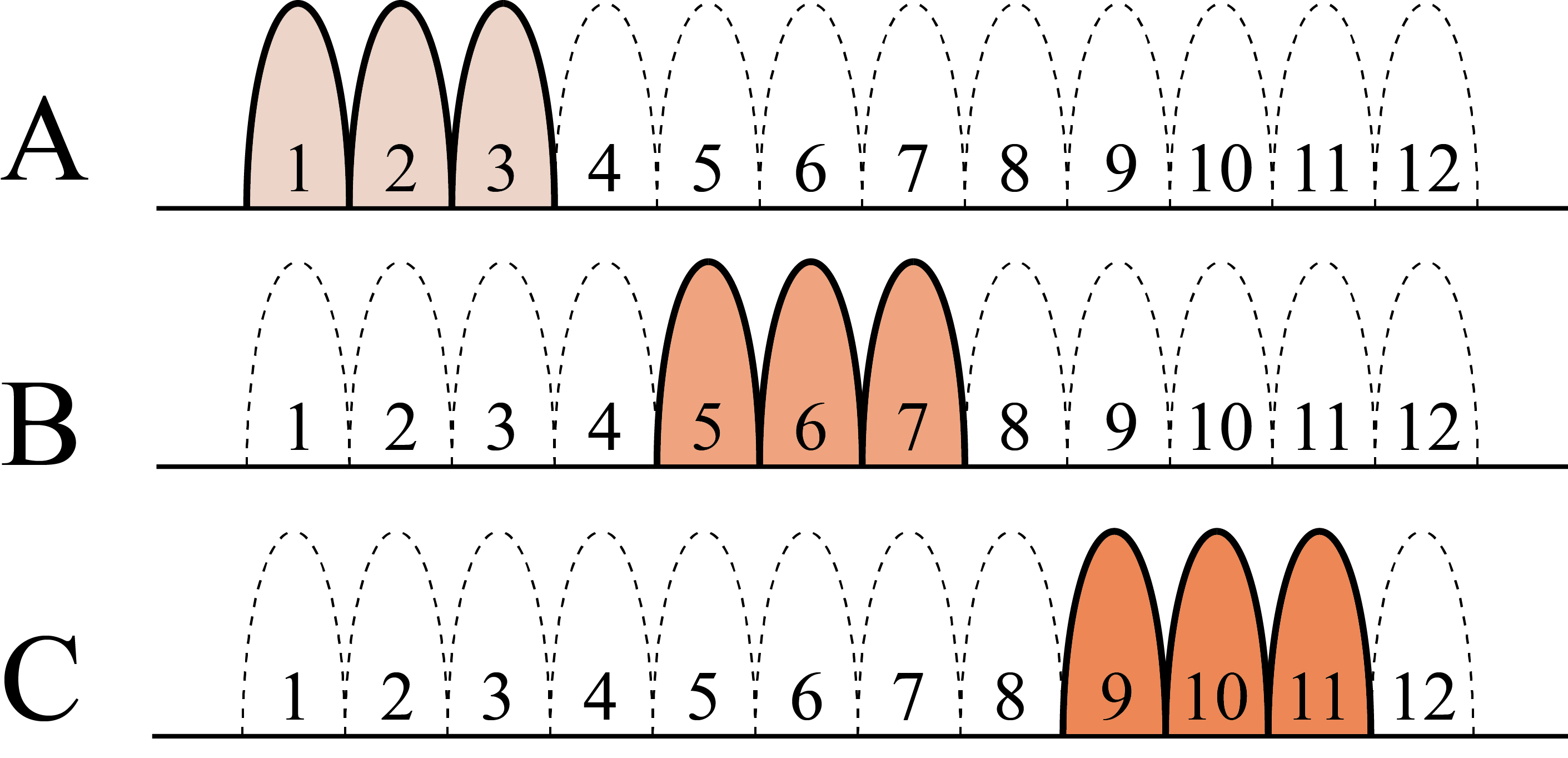}
    \caption{Channel used by interfering nodes A, B and C.}
  \label{fig:interfer}
\end{figure}
%%%%%%%%%%%%%%%%%%%%%%%%%%%%%%%%%%%%%%%%%%%%%

%%%%%%%%%%%%%%%%%%%%%%%%%%%%%%%%%%%%%%%%%%%%%
\begin{figure}[!t]
  \centering
    \includegraphics[width=0.4\textwidth]{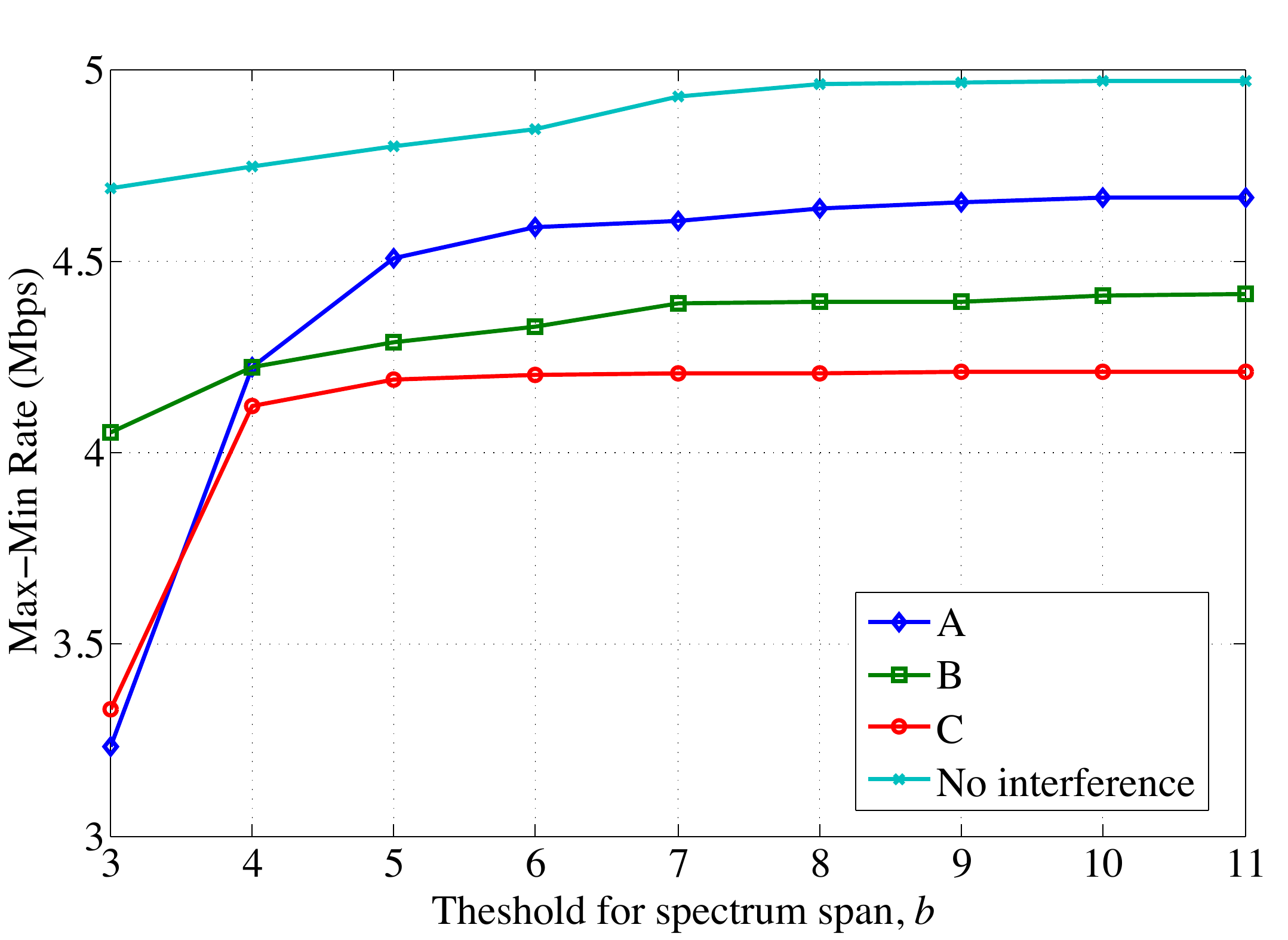}
    \caption{Max-min rate obtained for varying $b$ in presence of A, B or C.}
  \label{fig:span}
\end{figure}
%%%%%%%%%%%%%%%%%%%%%%%%%%%%%%%%%%%%%%%%%%%%%

The channel gain in each of the channels is generated using a Rician flat fading model with K-factor of 30dB. As shown in Fig. \ref{fig:interfer}, interfering nodes A, B and C operate in channels $(1,2,3)$, $(5, 6, 7)$ and $(9,10,11)$ respectively. These interfering nodes are transmitting at 33dB higher power than the noise power. We assume that these interfering nodes can be turned on or off independent of each other. 

As discussed in the previous section, there exists a trade-off between the max-min rate and threshold for spectrum span. The highest value of the max-min rate can be achieved when $b=M$. To get this trade-off curve, we use a brute-force algorithm written in C$++$ for multiple channel realizations.  For the simulation setup shown in Fig. \ref{fig:toposim} we obtain  this trade-off curve under the effect of interference from A, B and C, averaged over multiple channel realizations as shown in Fig. \ref{fig:span}. We observe that as the threshold for spectrum span increases, the max-min rate also increases. However we can see that in our simulations  as the threshold $b$ is increased to more than 5 channels, there is very little improvement ($2-5\%$) in max-min rate. This result indicates that the threshold $b$ can be chosen to be much smaller than $M$ while incurring only a small penalty on the max-min rate. 
For comparison purposes, we also solve the optimization problem in absence of interference from any of the interfering nodes and obtain the resulting  channel allocation and data rates.% for all nodes we introduce various interfering nodes. 
%Since , we first create a linear approximation of capacity constraint \cite{nazmul14powmin} and convert our problem to a mixed-integer linear program. 

\begin{figure}[!t]
  \centering
    \includegraphics[width=0.4\textwidth]{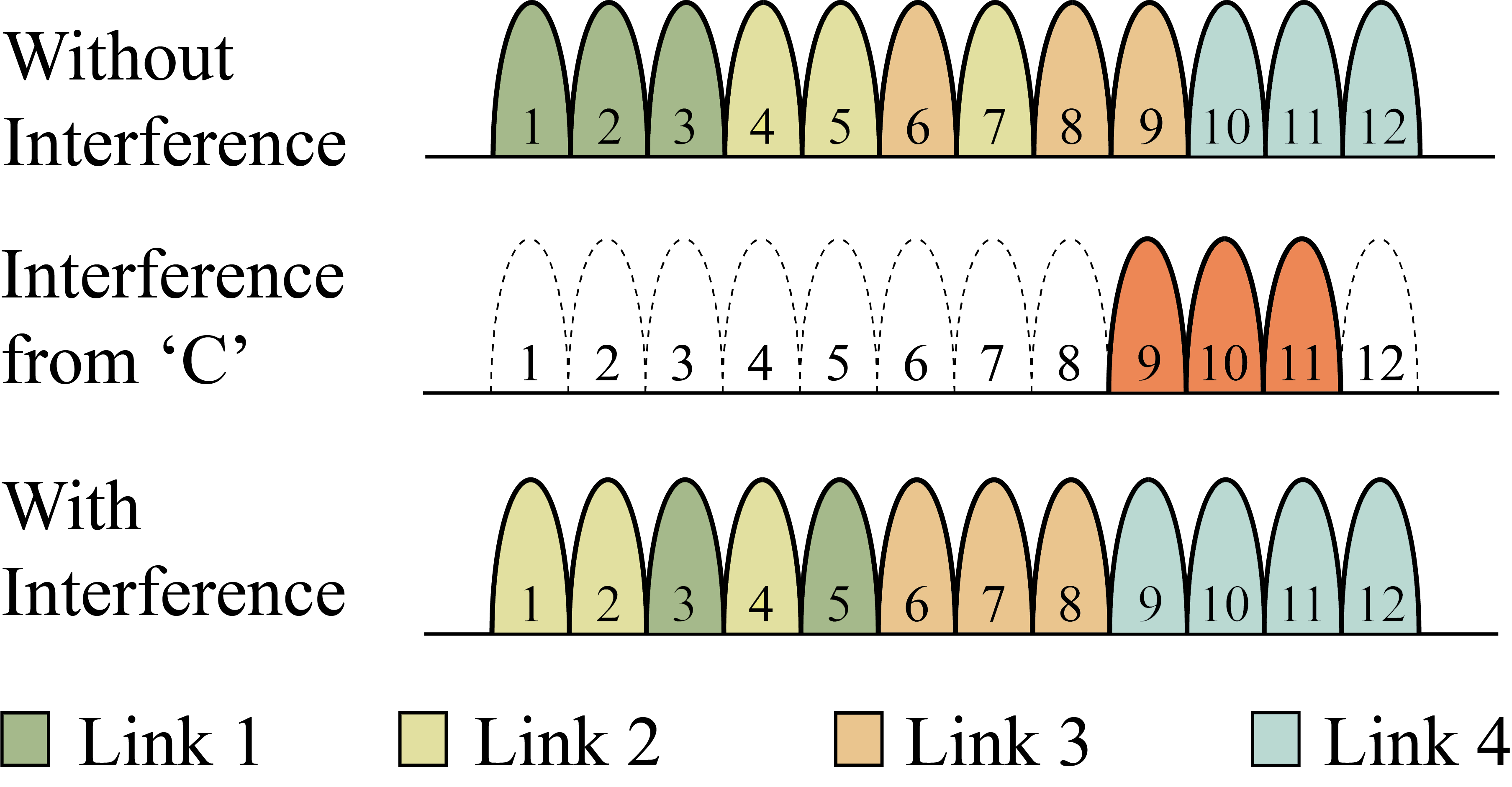}
    \caption{Change in channel allocation after introduction of interference from node `C' for $b=4$.}
  \label{fig:realloc}
\end{figure}

\begin{figure}[!t]
  \centering
    \includegraphics[width=0.37\textwidth]{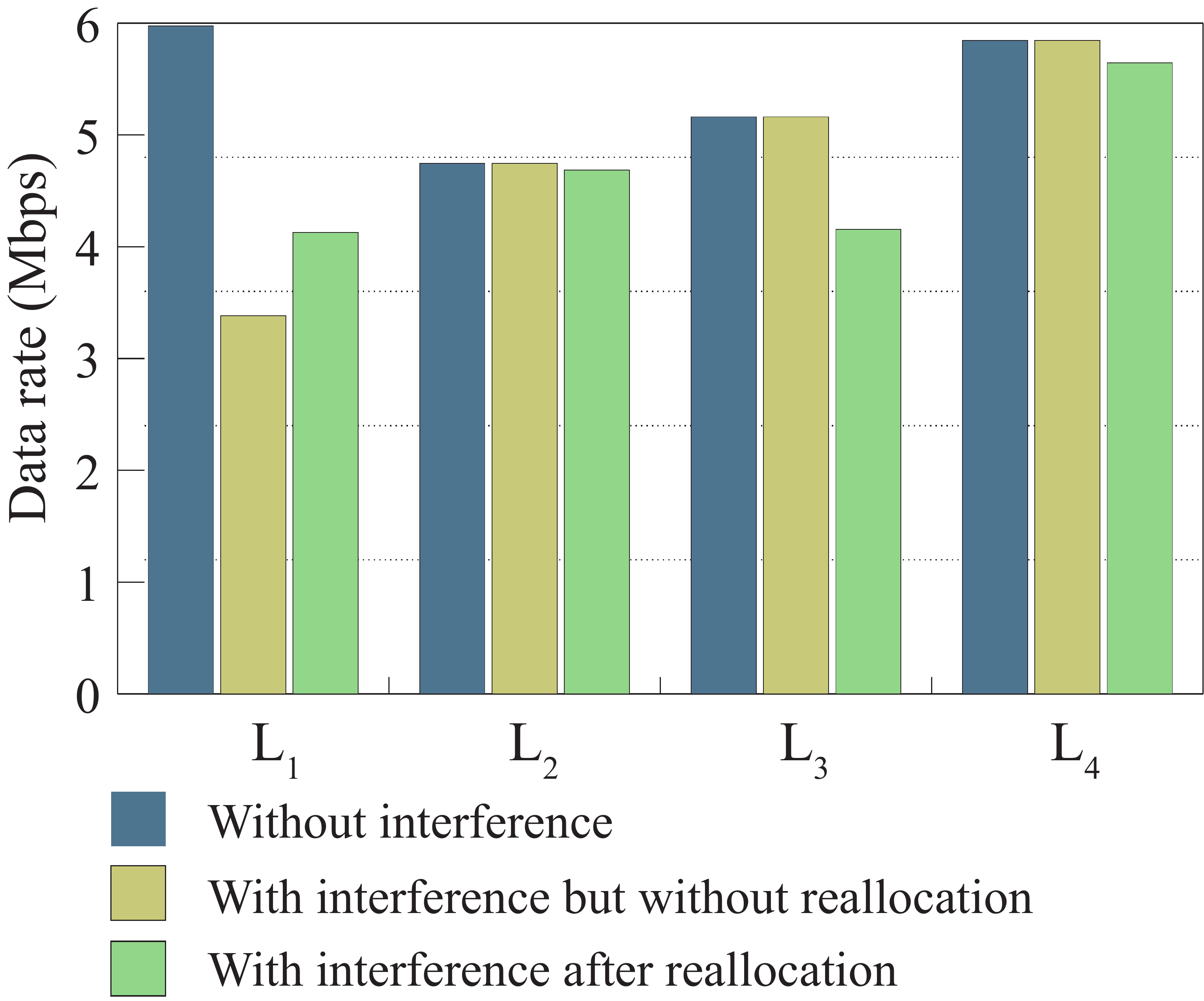}
    \caption{Data rate obtained by solving optimization problem with and without interference from node `A', $b=4$.}
  \label{fig:simbar}
\end{figure}

Fig. \ref{fig:realloc} shows the channel allocation for all links with and without interference from node C for a single channel realization with value of the spectrum span threshold, $b=4$. During this period nodes A and B are assumed to be turned off.  We can see that in the absence of interference, all links are allocated three channels each. The spectrum span for link $L_1$ and $L_4$ is 3 channels each while the spectrum span for link $L_2$ and $L_3$ is 4 channels each, under no effect of interference. In our simulation we find that the max-min rate is $4.74$ Mbps for this case. As the interference from node C is introduced, the performance of links $L_3$ and $L_4$ are degraded and the minimum rate among the links comes down to $3.65$ Mbps. However, if spectrum is reallocated based on our optimization formulation, this rate improves to $4.26$ Mbps. We can see that after the reallocation $L_3$ has moved out of interference and $L_4$ improves its performance by getting an extra channel. Fig. \ref{fig:simbar} shows the changes in max-min rate under the interference from A for all links. We see that interference from A affects only link $L_1$ and the minimum data rate among all links drops from $4.74$ Mbps to $3.38$ Mbps. However after the reallocation, the max-min rate improves to $4.2$ Mbps, and if we allow the value of $b$ to increase, we observe further improvements. 

%%%%%%%%%%%%%%%%%%%%%%%%%%%%%%%%%%%%%%%%%%%%%
%%%%%%%%%%%%%%%%%%%%%%%%%%%%%%%%%%%%%%%%%%%%%

%%%%%%%%%%%%%%%%%%%%%%%%%%%%%%%%%%%%%%%%%%%%%
%%%%%%%%%%%%%%%%%%%%%%%%%%%%%%%%%%%%%%%%%%%%%

%%%%%%%%%%%%%%%%%%%%%%%%%%%%%%%%%%%%%%%%%%%%%%%%%%%%%%%%%%%%%%%%%%%%%%%%%
\section{Experiments on orbit testbed}
\label{sec:exp}
\subsection{Platform}
We test a scaled down version of the above simulation with 4 USRP2 nodes using GNU Radio software platform on the ORBIT testbed. ORBIT testbed has a grid of 400 radio nodes in a $20\times 20 $ structure with $~ 1m$ distance between the nodes. Each of these 400 nodes is equipped with a variety of radio platforms including 802.11 a/b/g, Bluetooth, Zigbee and various versions of software defined radios such as USRP platforms, WARP platforms and the WINLAB developed CRKit cognitive radios \cite{orbit, ettus,murphy06warp}. In our experiments we use USRP N210, with SBX transceiver daughter-card with an operating spectral range of 400 MHz to 4.4 GHz and 100 mW transmit power.   GNU Radio is a free and open-source SDR framework and toolkit that provides the application programming interface (API) supporting, among other hardware platforms, range of USRP devices.  The NC-OFDM communication paths are implemented in C++ and Python in GNU Radio. Fig. \ref{fig:gnurblock} represents the block diagram of the transmitter and receiver in our experimental setup.  

%%%%%%%%%%%%%%%%%%%%%%%%%%%%%%%%%%%%%%%%%%%%%
\begin{figure}[!t]
  \centering
    \includegraphics[width=0.38\textwidth]{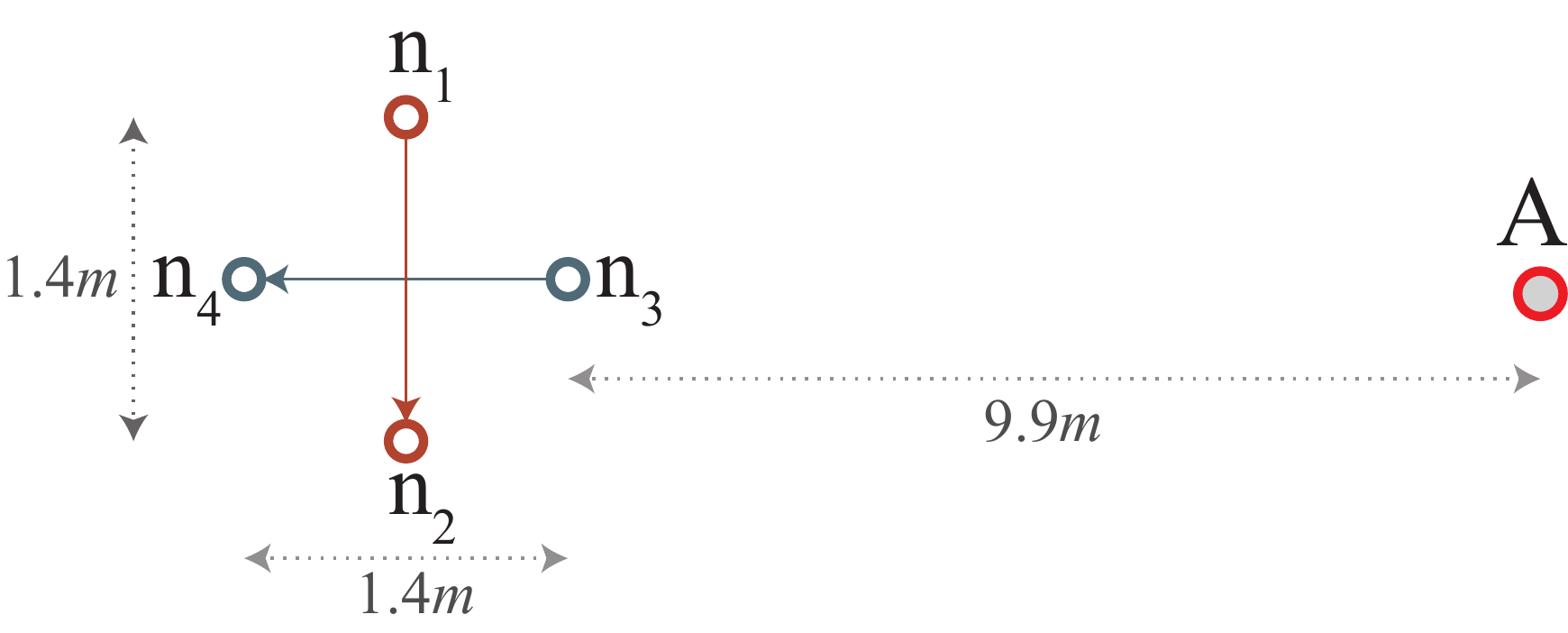}
    \caption{Topology used in the ORBIT testbed.}
  \label{fig:orbitsetup}
\end{figure}
%%%%%%%%%%%%%%%%%%%%%%%%%%%%%%%%%%%%%%%%%%%%%
 
%%%%%%%%%%%%%%%%%%%%%%%%%%%%%%%%%%%%%%%%%%%%%%%%%%%%%%%%%%%%%%%%%%%%%%%%%
\subsection{Experiments}
The experiments are designed for simultaneous communication between two links which use the parameters described in Table \ref{table:expparam}.  The topology shown in Fig. \ref{fig:orbitsetup} is used for the experiments on the ORBIT testbed.  Two point-to-point links are $L_1 = (n_1,n_2)$ and $L_2 = (n_3,n_4)$, and the node A is the interfering node. Among the available 128 subcarriers, we use only 112 subcarriers for data communication and use remaining 16 channels for  control and synchronization  purposes. Our experiments on this setup revealed that the OFDM implementations using the USRP2 platform are not robust when using less than 4 subcarriers on a single link. Therefore we group these 112 subcarriers into groups of 4 subcarriers to form 28 channels resulting in $31.25$KHz bandwidth for each channel. Therefore according to our system model
%%%%%%%%%%%%%%%%%%%%%%%%%%%%%%%%%%%%%%%%%%%%%
\begin{align}
  \mathcal{N} &= \{L_1,L_2\},\, \nonumber\\
  \mathcal{M} &= \{1,2,\ldots,27,28\}. \nonumber
\end{align}
%%%%%%%%%%%%%%%%%%%%%%%%%%%%%%%%%%%%%%%%%%%%%

%%%%%%%%%%%%%%%%%%%%%%%%%%%%%%%%%%%%%%%%%%%%%
\begin{figure}[!t]
  \centering
    \includegraphics[width=0.43\textwidth]{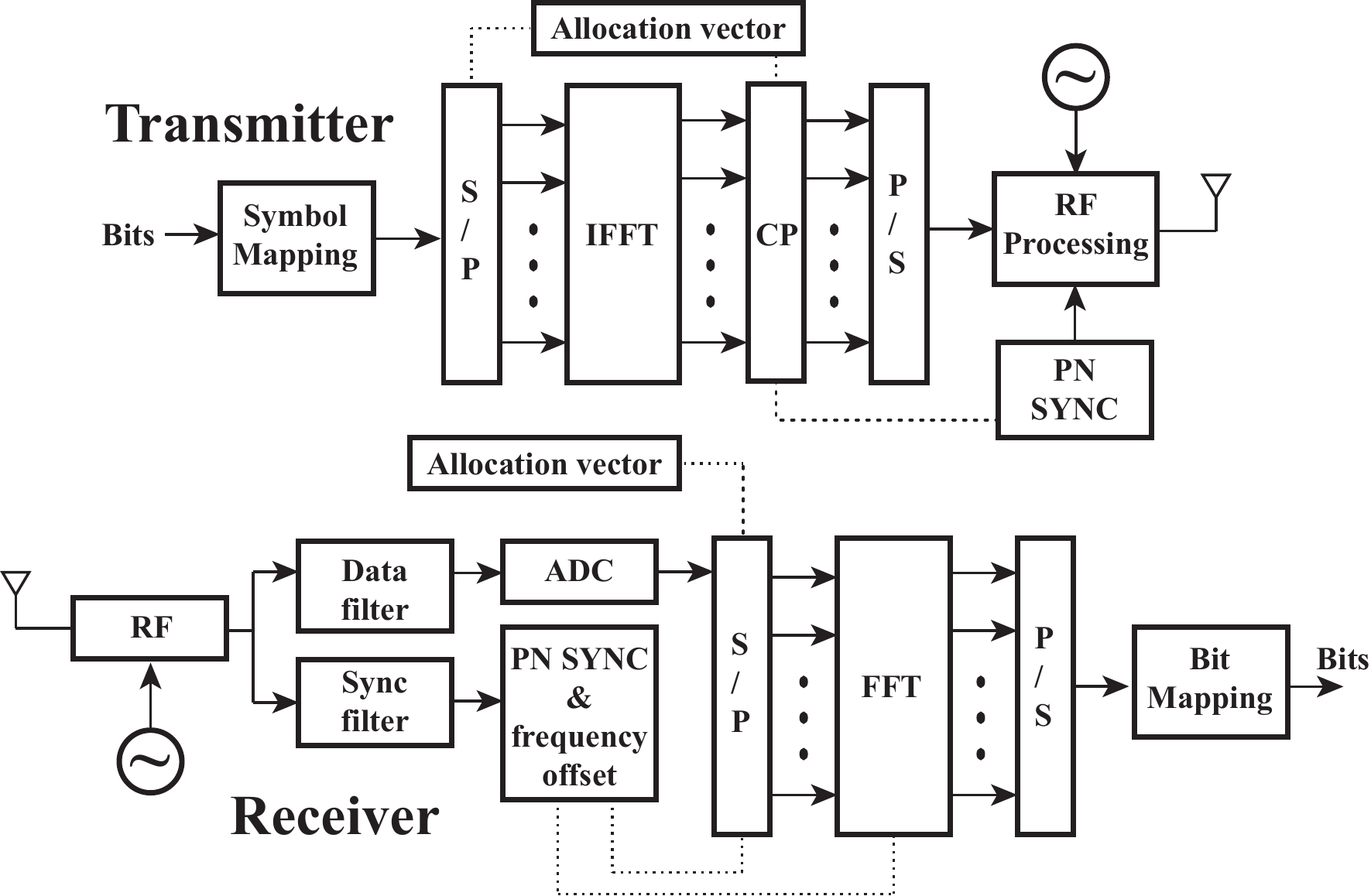}
    \caption{Block diagram for implementation of NC-OFDM with GNU Radio.}
  \label{fig:gnurblock}
\end{figure}
%%%%%%%%%%%%%%%%%%%%%%%%%%%%%%%%%%%%%%%%%%%%%

%%%%%%%%%%%%%%%%%%%%%%%%%%%%%%%%%%%%%%%%%%%%%
\begin{table}[t!]
\centering
\caption{Parameters used in experiment}
\begin{tabular}{|r|p{0.19\textwidth}|}
  \hline
  FFT Length & 128 \\
  \hline
  Size of data packet & 1500 bytes\\
  \hline
  Center frequency & 1.5GHz \\
  \hline
  Total available bandwidth & 1 MHz\\
  \hline
  Modulation type & BPSK \\
  \hline
\end{tabular}
\label{table:expparam}
\end{table}
%%%%%%%%%%%%%%%%%%%%%%%%%%%%%%%%%%%%%%%%%%%%%

One of the most important challenges faced during our experiments were related to synchronization and interference to a link from the sidelobe power of another links which are using the adjacent channels. Different methods of handling the issue of sidelobe power have been proposed such as, usage of guardband or sidelobe suppression \cite{dong2009,ghassemi2010,yuan2010}.  We note that the problem associated with synchronization  can also be addressed by using a filter with sharp cut-offs corresponding to the allocation vector at the transmitter and receiver. However designing these filters which are also reconfigurable is difficult and providing very sharp cut-off is very resource consuming. Instead,  to mitigate the  problem of synchronization, we implement an out-of-band synchronization method in our NC-OFDM implementation using PN-sequence preambles \cite{tufve99sync}.
%%%%%%%%%%%%%%%%%%%%%%%%%%%%%%%%%%%%%%%%%%%%%
\begin{figure}
%\begin{algorithm}
 \begin{boxedalgorithmic}[1]
 \renewcommand{\algorithmicrequire}{\textbf{Input:}}
 \renewcommand{\algorithmicensure}{\textbf{Output:}}
 \REQUIRE \textbf{A}, $r_l^m$, $r_{l}$
 \ENSURE  $\textbf{A}$
 %\\ \textit{Initialisation} :
  \STATE $l_b$ = link using channel $1$
  \STATE $m_b$ =  $1$
 %\\ \textit{LOOP Process}
  \FOR {$m = 2$ to $M$}
  \STATE $l_n$ = link using channel $m$
  \IF {($(l_n \ne l_b)$ and $(l_n \ne$ NULL))}
  \IF {($(r_{l_n} - r_{l_n}^m) \leq (r_{l_b} - r_{l_b}^{m_b})$)}
  \STATE $a_{{l_n}m}=0$,
  \STATE $r_{l_n}^m=0$,
  \ELSE
  \STATE $a_{{l_b}{m_b}}=0$,
  \STATE $r_{l_b}^{m_b}=0$,
  \ENDIF
  \ENDIF
  \STATE $m_b$ = $m$
  \STATE $l_b$ = $l_n$
  \ENDFOR
 \RETURN $A$ 
 \end{boxedalgorithmic} 
 \caption{Algorithm for creating guardband.}
 \label{fig:algogb}
 %\end{algorithm}
 \end{figure}
%%%%%%%%%%%%%%%%%%%%%%%%%%%%%%%%%%%%%%%%%%%%% 
Fig. \ref{fig:gnurblock} shows the block diagram of our implementation of NC-OFDM using GNU Radio. We modify the synchronization block to separate the  process of synchronization from data path. To address the problem of interference from sidelobe power, we provide a guardband between two adjacent channel allocations.

\begin{figure}[!t]
  \centering
    \includegraphics[width=0.37\textwidth]{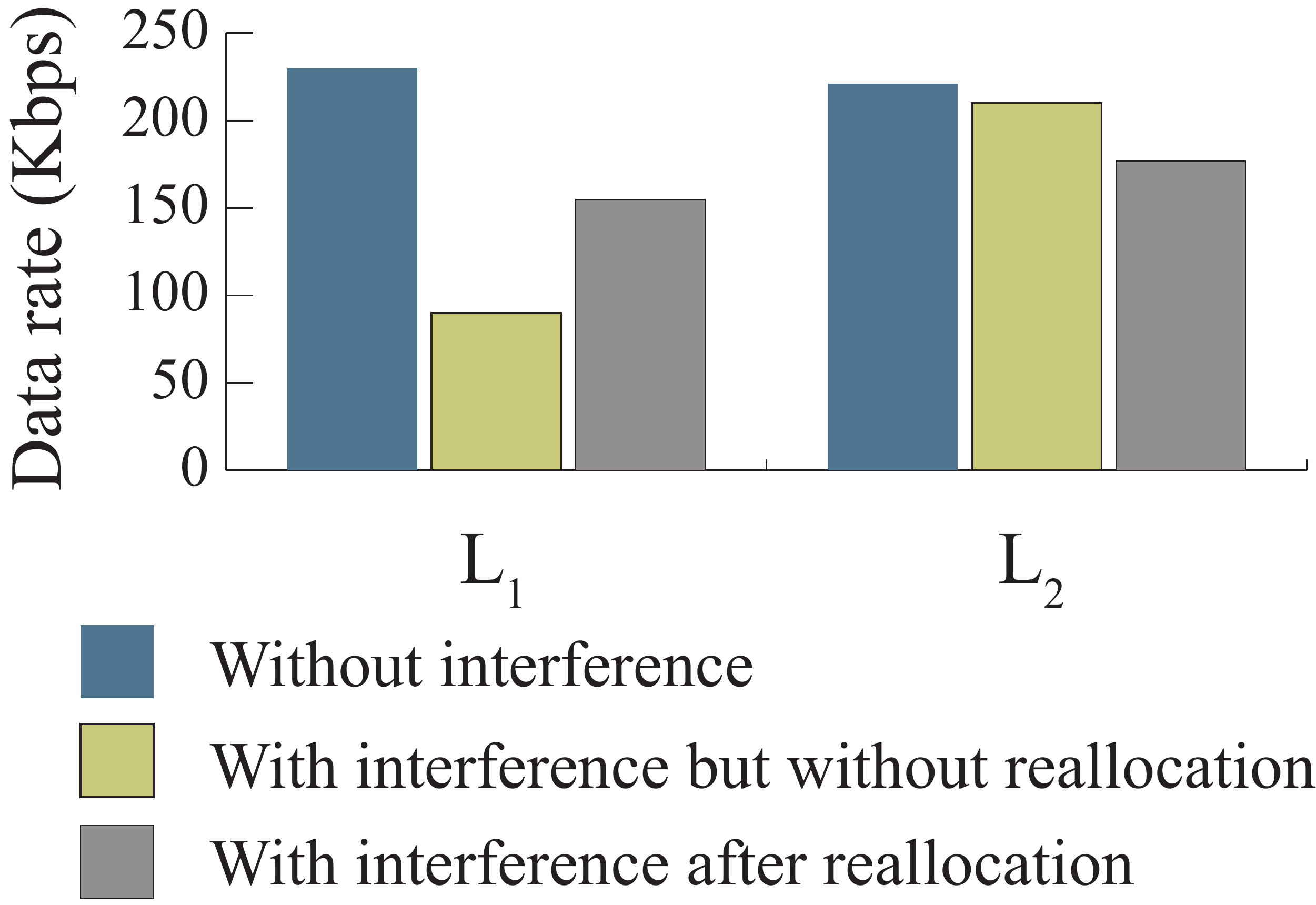}
    \caption{Data rate obtained in the ORBIT testbed.}
  \label{fig:orbresult}
\end{figure}

The channel allocation is decided by an outside node which behaves as a controller, and it allocates the channels under  different interference environment implementing the proposed  max-min rate optimization algorithm. In our experiments, the distance between USRPs is very small, therefore we approximate the channel between them using a line-of-sight path loss model to fit in our optimization problem. In our experiments, we find that a guardband of 4 subcarriers  was sufficient for successful communication. Once we get the allocation vector, we create these guardbands in such a way that it has minimal effect on the max-min rate obtained after optimization. We use  the algorithm presented in Fig. \ref{fig:algogb} to create the guardband for our experiments.  Fig. \ref{fig:orbresult} shows the changes in data rate for both links due to the  interference from node A averaged over 10 iterations. We find that even when we did not have the perfect knowledge about the channel gain, the trend in the variation of data rate is similar to that observed in our simulation results.

%%%%%%%%%%%%%%%%%%%%%%%%%%%%%%%%%%%%%%%%%%%%%
%%%%%%%%%%%%%%%%%%%%%%%%%%%%%%%%%%%%%%%%%%%%%

%%%%%%%%%%%%%%%%%%%%%%%%%%%%%%%%%%%%%%%%%%%%%%%%%%%%%%%%%%%%%%%%%%%%%%%%%
%%%%%%%%%%%%%%%%%%%%%%%%%%%%%%%%%%%%%%%%%%%%%%%%%%%%%%%%%%%%%%%%%%%%%%%%%
\section{Conclusion} 
\label{sec:conclude}
In this paper we considered the problem of spectrum allocation  that maximizes the minimum rate for NC-OFDM-enabled  point-to-point cognitive radio links under the spectrum span constraint. Under the constraint of constant transmit powers and orthogonal spectrum allocation, we formulated a mixed-integer linear program that can be solved efficiently using readily available solvers.  We showed that there exists a clear trade-off between the spectrum span of the spectrum allocation and max-min rate. We found that spectrum span can be kept to small value with very little penalty on the max-min rate under different interference scenarios.The max-min rate obtained from the spectrum allocation that is calculated using our optimization formulation showed improvement under different interference conditions. We also presented an experimental evaluation of the techniques developed in this paper using  USRP enabled ORBIT radio network testbed.  We found similar trends in the variation of  of data rate in our experiments to that observed in our simulation results.

\section*{Acknowledgment}
The authors thank Dr. Gokul Sridharan for several valuable discussions and insights that improved the presentation of the results in this paper.

\bibliographystyle{IEEEtran}
\bibliography{comsnets} \end{document}